\documentstyle[12pt]{article}
\input{epsf}

\title{\bf Gravonuclear Instabilities in
Post-Horizontal-Branch Stars}
\author{Allen V. Sweigart$^1$\\
John C. Lattanzio$^2$\\
James P. Gray$^2$\\
and Christopher A. Tout$^3$\\
\vspace{1cm}\\
\normalsize $^1$Laboratory for Astronomy and Solar Physics, NASA/Goddard Space Flight Center, USA\\
\normalsize $^2$Department of Mathematics and Statistics, Monash University, Australia\\
\normalsize $^3$Institute of Astronomy, Madingley Road, Cambridge, UK}
\date{\mbox{}}
\begin{document}
\maketitle
\pagestyle{empty}
%
%
\def\bull{\vrule height .9ex width .8ex depth -.1ex}
\makeatletter
\def\ps@plain{\let\@mkboth\gobbletwo
\def\@oddhead{}\def\@oddfoot{\hfil\tiny\bull\quad
``The Galactic Halo~: from Globular Clusters to Field Stars'';
35$^{\mbox{\rm th}}$ Li\`ege\ Int.\ Astroph.\ Coll., 1999\quad\bull}%
\def\@evenhead{}\let\@evenfoot\@oddfoot}
\makeatother
%
%
\def\beginrefer{\section*{References}%
\begin{quotation}\mbox{}\par}
\def\refer#1\par{{\setlength{\parindent}{-\leftmargin}\indent#1\par}}
\def\endrefer{\end{quotation}}
%
%
{\noindent\small{\bf Abstract:} 
We investigate the gravonuclear instabilities reported by Bono et
al. (1997a,b) during the onset
of helium-shell burning at the end of the horizontal-branch
(HB) phase.  These instabilities are characterized by
relaxation oscillations within the helium
shell which lead to loops in the evolutionary tracks.
We find the occurrence of these
instabilities depends critically on how
the breathing pulses are suppressed near the end of the HB phase.
If they are suppressed by omitting the
gravitational energy term in the stellar structure equations,
then the helium profile within the core at the end of the HB
phase will contain a broad region of varying helium abundance.
The helium-burning
shell which forms in this region is too thick to
be unstable, and gravonuclear instabilities do
not occur.  If, on the other hand, the breathing pulses are
suppressed by prohibiting any increase in the central helium
abundance, then the final helium profile can exhibit a large
discontinuity at the edge of the helium-exhausted
core.  The helium shell which forms just
exterior to this discontinuity is then much thinner and
can be thermally unstable.  Even in this case, however,
the gravonuclear instabilities disappear as soon as the nuclear
burning broadens the helium shell into its characteristic S-shape.
We conclude that the gravonuclear instabilities found
by Bono et al. are a  consequence of the ad hoc procedure
used to suppress the breathing pulses.}
%
%
\section{Introduction}
A number of intriguing astrophysical problems are associated
with the termination of central helium burning in
horizontal-branch (HB) stars.  One example
concerns the so-called ``breathing pulses'' of the
convective core.  While numerical algorithms for treating
semiconvection (e.g., Robertson \& Faulkner 1972) 
work quite effectively during most of the
HB phase, they invariably fail once the central
helium abundance $Y_c$ falls below $\simeq 0.1$.  
At that time the
convective core suddenly grows so large that
it engulfs most of the previous semiconvective
zone, 
thereby bringing so much fresh helium into the center that $Y_c$ increases.

These breathing pulses are generally suppressed
using one of the following methods:

$\bullet$ {\sl Omission of gravitational energy term $\epsilon_g$.}
Dorman \& Rood (1993) have shown that the
breathing pulses can be suppressed by setting $\epsilon_g$ equal to 0
in the stellar structure equations during the core-helium-exhaustion phase. 
 An example
of a canonical HB and post-HB track computed with this approach is given in
Figure 1a.  Note that the evolution is quite smooth
without any indication of an instability.  The helium-burning
luminosity $L_{He}$ along this sequence, given in Figure 2a, shows only
a characteristic dip at the end of the HB phase, as the helium burning shifts
outward from the center to a shell.  Most
importantly, the composition profile
within the core at the end of the HB phase contains
a broad region of varying helium abundance $Y$
corresponding to the former semiconvective
zone (see Figure 3a).

\begin{figure}
\centerline{\epsfxsize=0.8\hsize \epsffile{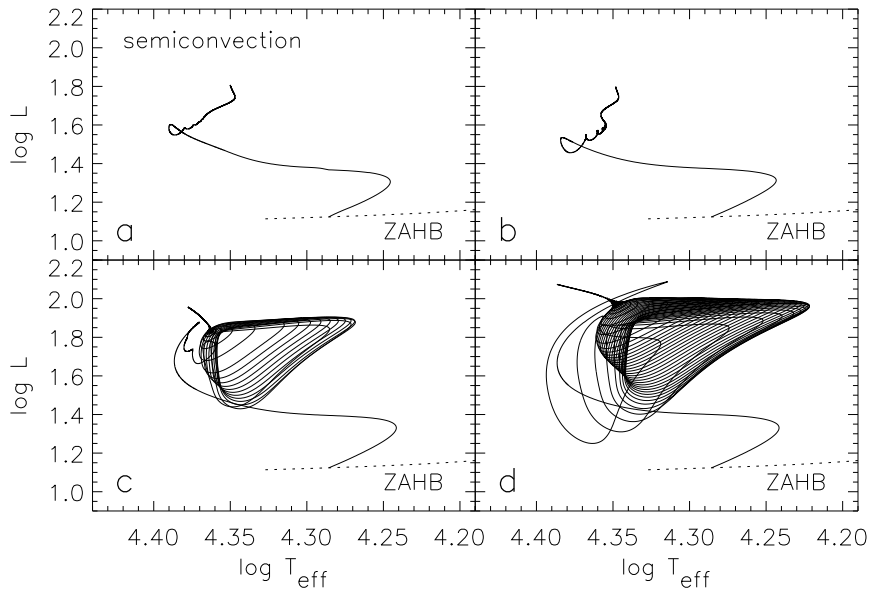}\hskip 1.5 truecm}
\caption{Evolutionary tracks for the HB and post-HB phases
of a $0.48M_\odot$ star with 
$Z = 0.03$.  Panel (a) gives a canonical semiconvection track
computed with the $\epsilon_g  = 0$ method for suppressing
the breathing pulses during the core-helium-exhaustion phase.  The
tracks in panels (b), (c) and (d) were computed with the composition
algorithm described in the text (see \S 2).}
\end{figure}

$\bullet$ {\sl Limit on growth of the convective core.} 
An alternative method for suppressing the breathing pulses,
used by Bono et al. (1997a,b), is
to limit the rate at which the convective
core can grow in order to prevent $Y_c$ from increasing.  Inspection of
Tables 1 -- 5 of Castellani et al. (1991) shows that this ad hoc
method leads to a greatly enlarged convective core that
remains large until helium exhaustion.  We infer therefore that the final
helium profile should then contain a large
discontinuity at the edge of the helium-exhausted core
and thus should be markedly different from the
profile produced by the first suppression method.

Bono et al. (1997a,b) recently argued that the onset of
helium-shell burning in post-HB stars is dramatically different
from the smooth evolution shown in Figure 1a, especially
for metal-rich stars with low envelope masses.  Their
results indicate that the helium-burning shell undergoes
a series of ``gravonuclear instabilities'' caused
by relaxation oscillations similar
to the helium-shell flashes that occur later on the asymptotic
giant branch.
These
instabilities lead to pronounced ``gravonuclear loops'' (GNLs)
along the evolutionary tracks, which could have interesting
observational consequences.  Figures 1d and 2d
give examples of these GNLs and the associated helium-shell
instabilities (see \S 2).
We have undertaken extensive  calculations to understand
the cause of these gravonuclear instabilities.  Our
main conclusions are:

{\sl The occurrence of gravonuclear instabilities depends critically
on the helium profile within the core  at the end of the
HB phase and hence on the method used to suppress
the breathing pulses.  Gravonuclear instabilities
are only found when there is a large discontinuity in the
helium abundance, which forces the helium burning to be
confined to a narrow region at the edge of the core.  
Contrary to the Bono et al.
results, we find that gravonuclear instabilities are not
caused by a high envelope opacity nor do they depend on the
envelope mass.  Rather, they are a  consequence of the
method used by Bono et al. to suppress the breathing
pulses.}

\section{Dependence of Instabilities on Core-Helium Profile}
We first explored the dependence of the gravonuclear instabilities
on the composition profile within the core at the end of the
HB phase.  To do this, we computed a number of HB and post-HB evolutionary
sequences for a star with a mass $M  =  0.48 M_\odot$ and
a heavy-element abundance $Z  =  0.03$
for various assumptions about the final composition
profile.  These model parameters were chosen to optimize the
likelihood of gravonuclear instabilities according to
the Bono et al. results.  A very small time step of only
400 yr was used in the post-HB models in order to resolve any
instabilities, if present.  Moreover,
a thermal stability analysis was performed on each of the $\simeq  30,000$
post-HB models in each sequence to search for any
unstable modes.

\begin{figure}
\centerline{\epsfxsize=0.8\hsize \epsffile{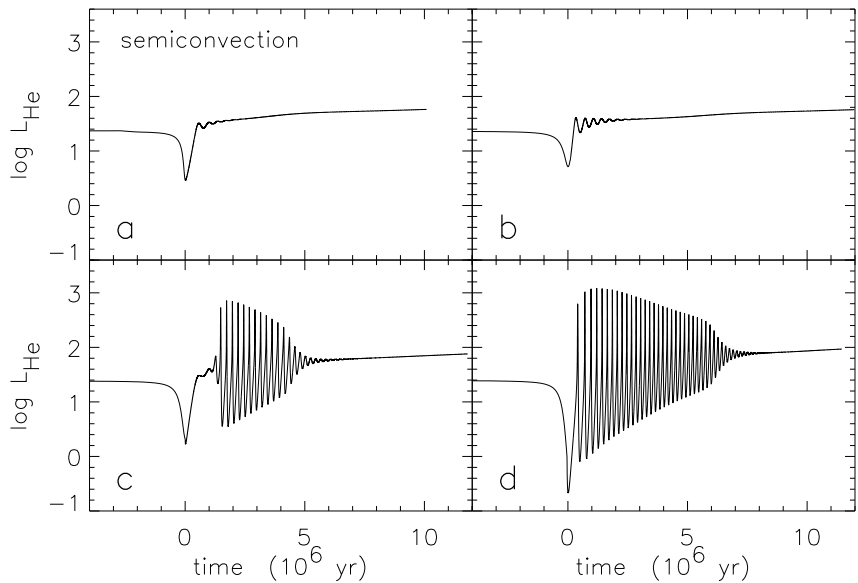}\hskip 1.5 truecm}
\caption{Time dependence of the helium-burning luminosity
$L_{He}$ during the post-HB evolution of the tracks
in Figure 1.  The zero point of the timescale corresponds
to the minimum in $L_{He}$ at the end of the HB
phase.  Each panel refers to the corresponding panel in Figure 1.}
\end{figure}

\begin{figure}
\centerline{\epsfxsize=0.8\hsize \epsffile{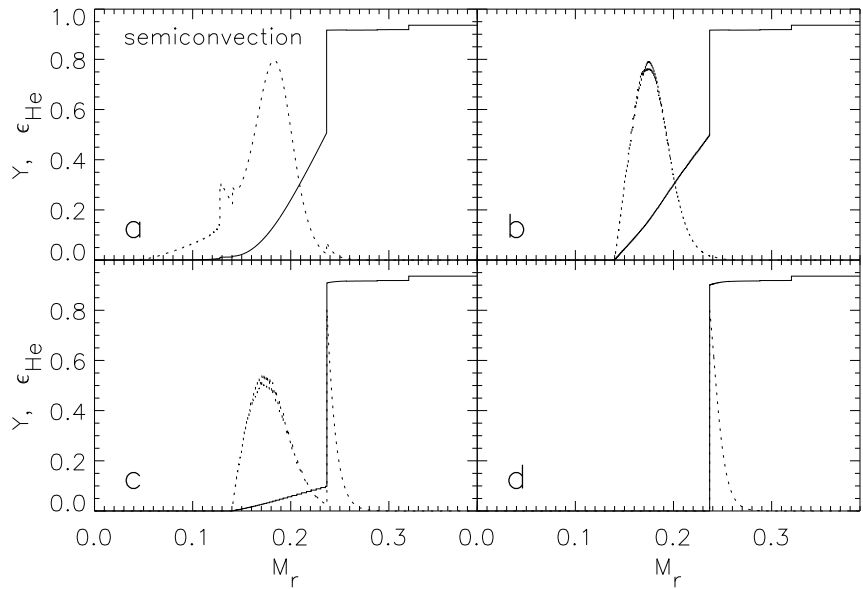}\hskip 1.5 truecm}
\caption{Helium profiles at central helium exhaustion
for the sequences in Figures 1 and
2 (solid curves).  The mass $M_r$ is in
solar units.  The helium-burning rate
$\epsilon_{He}$ (arbitrary units) is shown by dotted
curves.  Panels (a)--(d)  correspond to the panels in
Figures 1 and 2.}
\end{figure}

Our first sequence was a canonical semiconvection sequence computed
with the $\epsilon_g  = 0$ method for suppressing the
breathing pulses.  The results, given in Figures 1a and 2a,
show no signs of any instability.  In particular, the onset of
helium-shell burning is marked by only minor (and damped) ringing
in the helium-burning luminosity.  The composition profile
for this sequence, given in Figure 3a, contains a
discontinuity in the helium abundance at $M_r = 0.237 M_\odot$
corresponding to the outer edge of the semiconvective zone at its
maximum extent during the HB phase.  Interior to this
discontinuity there is a broad region of varying helium abundance
which also is a remnant of the previous semiconvection
and which we will refer to as the ``helium tail''.  Note
that the helium burning in Figure 3a covers a
wide range in mass.

We repeated these calculations using a ``composition
algorithm'' to specify the size of the helium-depleted region
instead of the canonical semiconvection
algorithm.  Essentially we required that the composition
be completely mixed from the center out to a specified
mass point in each model regardless of whether this region
was fully convective.  No mixing was permitted outside this
point.  By varying the size of this
mixed region during the HB evolution we were able to generate different composition
profiles having more or less steep helium tails at
core-helium exhaustion.  All these profiles had a helium discontinuity
at $M_r = 0.237  M_\odot$, as found with canonical
semiconvection.  This composition
algorithm was turned off at the end of the HB,
and  post-HB evolution was then followed in the
same manner as the canonical case.

Figures 1b, 2b and 3b present the results
for a sequence computed with this 
algorithm. The final composition profile
given in Figure 3b was chosen to mimic
the composition profile for the semiconvection
sequence given in Figure 3a.  The
resulting track morphology and helium-burning luminosity
are virtually identical to those for the semiconvection
sequence.  Moreover, no thermally unstable modes were
found.  This indicates that the post-HB evolution is
not sensitive to how the composition profile at
the end of the HB phase is actually produced.

We then used our composition algorithm to compute
sequences with shallower helium tails.  No gravonuclear instabilities
were found until the size of the helium tail was reduced to
that shown in Figure 3c.  The helium
burning in Figure 3c initially covered a broad region
within the helium tail, and the models were
then stable.  After $\simeq 10^6$ yr following
core-helium exhaustion, however, this helium tail
burned away, and the helium burning then shifted outward
to a narrow region just outside the helium
discontinuity.  The helium burning immediately became
unstable, giving rise to the flashes in Figure 2c
and the GNLs in Figure 1c.  The stability
analysis of these models revealed the existence
of thermally unstable modes.

We have also considered the limiting case of
a helium discontinuity and no tail
(Figure 3d).  As shown in
Figures 1d and 2d, such a
profile leads to strong gravonuclear instabilities
and to large GNLs.
The composition profile in Figure 3d should
be similar to the profile produced by the Bono
et al. method for suppressing the breathing
pulses.  Note that this method forces the helium
burning to be confined to a narrow region just outside
the helium discontinuity.

The above results are not surprising.  Schwarzschild \&
H\"arm (1965) showed that a nuclear
burning shell must be thin to be
unstable.  Figures 1, 2 and 3 confirm that
gravonuclear instabilities do not occur when the
helium-burning region is broad.  Only when the composition
profile confines the burning to a narrow region,
as in Figure 3d, do we find gravonuclear instabilities.

As further confirmation of this result, 
Figure 4 shows the helium profile at four times
during the post-HB evolution of the sequence plotted
in Figures 1d, 2d and 3d.  Panel (a) 
is the same as Figure 3d and
corresponds to the onset of gravonuclear instability.  Panel (b)
shows the helium profile during the
period of strong gravonuclear instability, while panel (c)
shows the helium profile for the last model in
which we found thermally unstable modes.  The
helium-burning region in Figure 4 progressively
broadens with time until by panel (d) the models are
completely stable.  We conclude therefore that the
gravonuclear instabilities disappear as soon as
the helium-burning region broadens into its characteristic
S-shape profile.

\begin{figure}
\centerline{\epsfxsize=0.8\hsize \epsffile{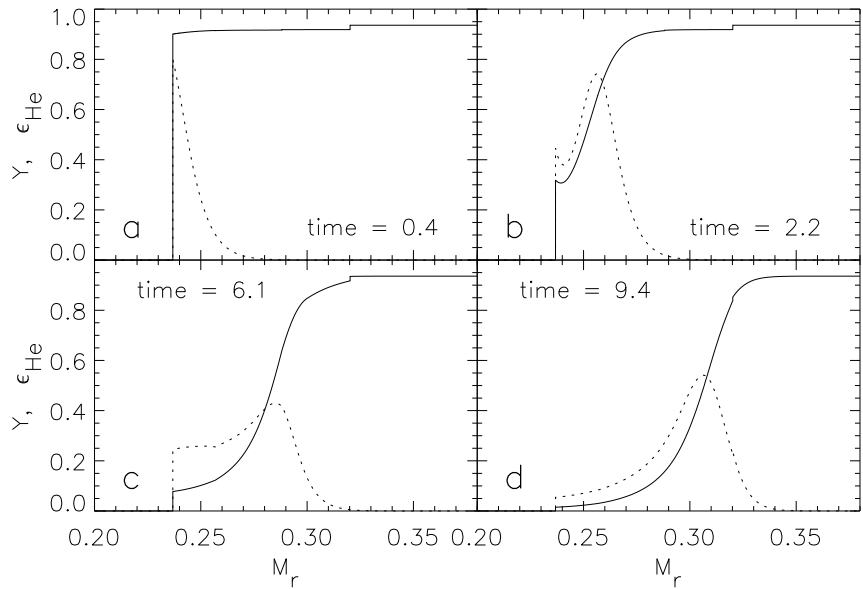}\hskip 1.5 truecm}
\caption{Helium profiles at various
times during the post-HB evolution of the sequence in Figure 2d 
(solid curves).  The helium-burning rate
$\epsilon_{He}$ is shown in arbitrary units by the dotted curves.  The times
in each panel refer to the timescale in 
Figure 2d in units of $10^6$ yr.}
\end{figure}

\section{Dependence of Instabilities on Envelope Mass and Z}
We have also investigated whether the gravonuclear instabilities
depend on the envelope mass and metallicity, as suggested by
Bono et al. Figure 5 shows the time dependence of $L_{He}$
during the post-HB evolution of four sequences with
$M  = 0.70  M_\odot$ (and hence larger envelope
mass).  The composition profiles at the end of the HB phase
for these sequences are virtually identical to
those in the corresponding panels of Figure 3. 
Figure 5 shows the same overall
behavior of $L_{He}$ as Figure~2, except for the shorter timescale
of the instabilities caused by the higher $L_{He}$
of these higher mass sequences.
It is clear therefore that the occurrence
of gravonuclear instabilities does not depend on
the envelope mass. The same conclusion applies to the metallicity.  
Calculations for a
$M = 0.52 M_\odot$, $Z = 0.002$ sequence, computed for the same helium profile
as in Figure 3d, show extensive GNLs just as for our higher metallicity
sequences.

\begin{figure}
\centerline{\epsfxsize=0.8\hsize \epsffile{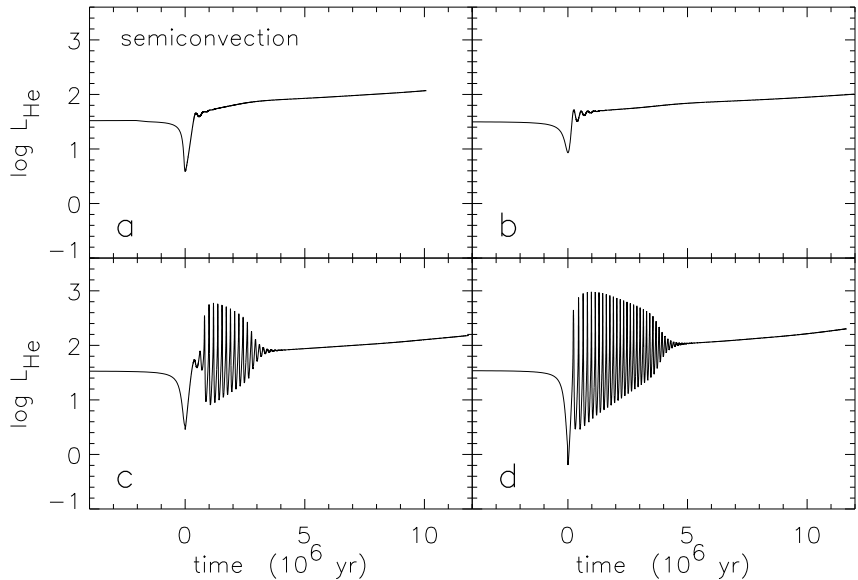}\hskip 1.5 truecm}
\caption{Same as Figure 2 except for a $0.70  M_\odot$ star.}
\end{figure}

\section{Conclusions}
Extensive calculations obtained with three independent
stellar evolution codes have enabled us to consistently
induce or remove GNLs according to the helium-composition
profile at the end of core-helium burning.  We
have shown that GNLs are not favored by higher metallicities
or lower masses, as postulated by Bono et al. (1997a,b).  Rather,
GNLs are produced by the narrowness of the helium-burning region
when the helium-burning shell ignites immediately following
core-helium exhaustion.  This is most influenced
by the way that the convective and semiconvective regions
are calculated at the completion of core-helium burning and
further illustrates our incomplete knowledge of this
complicated phase.  If we could observationally determine
the existence, or otherwise, of GNLs, it would be a direct
probe into the helium profile at
helium exhaustion and hence provide information about the
occurrence of core-breathing pulses.

\beginrefer

\refer Bono, G., Caputo, F., Cassisi, S., Castellani, V. \& Marconi, M. 1997a,
ApJ, 479, 279

\refer Bono, G., Caputo, F., Cassisi, S., Castellani, V. \& Marconi, M. 1997b,
ApJ, 489, 822

\refer Castellani, V., Chieffi, A., \& Pulone, L. 1991, ApJS, 76, 911

\refer Dorman, B., \& Rood, R. T. 1993, ApJ, 409, 387

\refer Robertson, J. W., \& Faulkner, D. J. 1972, ApJ, 171, 309

\refer Schwarzschild, M., \& H\"arm, R. 1965, ApJ, 142, 855

\endrefer

\end{document}